\documentclass[pra,twocolumn,showpacs,preprintnumbers,superscriptaddress]
{revtex4}

\usepackage{times}
\usepackage{bm}
\usepackage{float}
\usepackage{graphicx}
\usepackage{amsbsy}
\usepackage{amsmath,mathtools}
\usepackage{amsfonts}
\usepackage{amsthm}
\usepackage{xcolor}

\begin{document}

\theoremstyle{plain}
\newtheorem{theorem}{Theorem}
\newtheorem{lemma}[theorem]{Lemma}
\newtheorem{corollary}[theorem]{Corollary}
\newtheorem{proposition}[theorem]{Proposition}
\newtheorem{conjecture}[theorem]{Conjecture}

\theoremstyle{definition}
\newtheorem{definition}[theorem]{Definition}

\theoremstyle{remark}
\newtheorem*{remark}{Remark}
\newtheorem{example}{Example}
\title{Witness Operator Provides Better Estimate of the Lower Bound of Concurrence of Bipartite Bound Entangled States in $d_{1}\otimes d_{2}$ Dimensional System}
\author{Shruti Aggarwal, Satyabrata Adhikari}
\email{shruti_phd2k19@dtu.ac.in, satyabrata@dtu.ac.in} \affiliation{Delhi Technological
University, Delhi-110042, Delhi, India}

\begin{abstract}
It is known that the witness operator is useful in the detection and quantification of entangled states. This motivated us for the construction of the family of witness operators that can detect many mixed entangled states. This family of witness operators is then used to estimate the lower bound of concurrence of the detected mixed entangled states. Our method of construction of witness operator is important in the sense that it will estimate a better lower bound of concurrence of the entangled states in arbitrary $d_{1}\otimes d_{2} (d_{1}\leq d_{2})$ dimensional system compared to the lower bound of the concurrence given in \cite{kchen}. We have shown the significance of our constructed witness operator by detecting many bound entangled states that are not detected by the earlier methods and then we use the expectation value of the witness operator to estimate the lower bound of the concurrence of those bound entangled states.
\end{abstract}
 \maketitle

\section{Introduction}
Entanglement \cite{schrodinger} is the essence of quantum formalism and can be considered as the heart of quantum information science. It is a fully quantum mechanical phenomenon that has no classical analogue. We can find quantum advantage over classical if we use entangled states in quantum information tasks such as quantum teleportation \cite{bennett}, superdense coding \cite{bennett1}, remote state preparation \cite{akpati}, quantum cryptography \cite{gisin} and quantum computation \cite{ekert}. The detection of entangled states is an important problem for the following reasons: (i) When an experiment is particularly carried out to generate a bipartite or multipartite entangled state, it is a challenging task to verify whether the generated state is entangled. (ii) Entanglement \cite{rhorodecki} serves as a useful ingredient in many applications of quantum information. Another important theoretical challenge in the theory of quantum entanglement is to give a proper description and quantification of quantum entanglement of bipartite and multipartite quantum systems. \\
Despite many efforts in the last decade, a completely satisfactory solution to both the problems has not been found. Attending to the first one the so-called separability problem, there exist, nevertheless, several sufficient conditions for the detection of entanglement. The first criterion to detect the entanglement was proposed by Peres \cite{peres} and it is known as partial transposition criterion. Later, it was proved to be a necessary and sufficient criterion for $2\otimes 2$ and $2\otimes 3$ systems \cite{mhorodecki}. The partial transposition criterion tells us that if the partial transposition of the bipartite state described by the density operator $\rho$ in $d_{1} \otimes d_{2}$ dimensional system has at least one negative eigenvalue then the state $\rho$ is said to be the negative partial transpose entangled state (NPTES). The $d_{1}\otimes d_{2}$ dimensional NPTES of rank at most $max(d_{1},d_{2})$ have been shown to be distillable entangled states \cite{phorodecki,chen}. Apart from NPTES, there exist other entangled states in the $d_{1}\otimes d_{2}$ $(d_{1},d_{2}\geq 3)$ dimensional system that are not detected by partial transposition criterion, i.e., these states are described by positive semidefinite matrix even after the application of partial transposition operation on them. This kind of entangled states are known as positive partial transpose entangled states (PPTES). PPTES are commonly called bound entangled states (BES). BES are very weak entangled states and they are not distillable by performing local operation and classical communication (LOCC). Divincenzo et.al. \cite{divincenzo} have provided the evidence for the existence of bound entangled states with negative partial transpose. Partial transposition method fails to identify BES so we require other methods that can detect BES. The methods to detect BES that we will use in this work are: (i) Realignment criterion \cite{rudolph}  and (ii) Witness operator \cite{mhorodecki,mlewenstein}.\\

\textit{Realignment criterion.}--- Let us consider a bipartite state described by the density operator $\rho=\sum_{ijkl}\rho_{ij,kl}|ij\rangle\langle kl|$ in ${\mathcal{H}}^{d_1}_A \otimes {\mathcal{H}}^{d_2}_B$ where ${\mathcal{H}}^{d_1}_A$ and ${\mathcal{H}}^{d_2}_B$ are Hilbert spaces of dimension $d_1$ and $d_2$, for two systems $A$ and $B$, respectively. Let $R$ be the realignment operation and after applying it on $\rho$, the output will take the form as $\rho^{R}=\sum_{ijkl}\rho_{kj,il}|kj\rangle\langle il|$. If the state $\rho$ is separable then $\|\rho^{R}\|_{1}\leq 1$ holds, where $\|.\|_{1}$ denotes the trace norm and defined by $\|H\|_1=Tr(\sqrt{HH^{\dagger}})$. We note here an important fact that since the eigenvalues of $\rho^{R}$ may be complex also, so the operator $\rho^{R}$ may not be a hermitian operator. But in this work, we have studied the examples in which we find that the operator $(\rho^{R})^{T_{B}}$ has real eigenvalues, where $(.)^{T_{B}}$ denotes the partial transposition with respect to the subsystem $B$.\\
\textit{Witness operator.}---It is a hermitian operator that separates the entangled states from separable states \cite{mhorodecki,terhal1}. It is an observable with at least one negative eigenvalue. Mathematically, $W$ is said to be an entanglement witness operator if
\begin{eqnarray}
&&(i) Tr(W\sigma)\geq 0,~~ \textrm{for all separable state}~~ \sigma~~ \textrm{ and}
\nonumber\\&& (ii) Tr(W\rho)< 0,~~\textrm{for at least one entangled state}~\rho\nonumber
\label{defwit}
\end{eqnarray}
There are two classes of witness operator: Decomposable witness operators and indecomposable witness operators. Former detects only NPTES and the later class of witness operators detect BES together with NPTES. Terhal \cite{terhal1} first introduced a family of indecomposable positive linear maps based on entangled quantum states using the notion of unextendible product basis. Soon after this work, Lewenstein et. al. \cite{mlewenstein} extensively studied the indecomposable witness operator and provided an algorithm to optimize them. The construction of witness operators are important in the sense that they can be used in an experimental set up to detect whether the generated state in an experiment is entangled. There are different methods of the construction of witness operator in the literature \cite{ganguly,adhikari1,sarbicki,halder,wiesniak}.\\
Witness operators can also be used in the detection of entangled states that act as a resource state in the teleportation protocol \cite{ganguly1,adhikari2}. Witness operator is also used in the quantification of entanglement \cite{brandao} and in the estimation of lower bound of the concurrence of the entangled states in $d_{1}\otimes d_{2} (d_{1}\leq d_{2})$ dimensional systems \cite{mintert2,kchen}.\\
The motivation of this work is two fold. Firstly, since BES are very weak entangled states so they behave like separable states and thus it is very difficult to separate BES from the set of separable states. Hence, in particular, the detection of BES is an important problem to consider. Further, we find application of bipartite BES in quantum cryptography \cite{phorodecki1} and thus by realizing its importance, we have constructed the witness operator for the detection of BES. Secondly, it is known that for higher dimensional systems, we don't have any closed formula for concurrence, just like we have for a two-qubit system. Thus, the quantification of entanglement by estimating the exact value of the concurrence is a formidable task. In spite of these, few attempts have been made to obtain lower bound of the concurrence for a qubit-qudit system \cite{gerjuoy,lozinski} and to derive purely algebraic lower bound of the concurrence \cite{fmintert}. Later, Chen et.al. \cite{kchen} have derived the lower bound of the concurrence for arbitrary $d_{1}\otimes d_{2}$ ($d_{1} \leq d_{2}$) dimensional system and it is given by
\begin{eqnarray}
C(\rho)\geq \sqrt{\frac{2}{d_{1}(d_{1}-1)}}\left(max(\|\rho^{T_{A}}\|_1,\|\rho^{R}\|_1)-1\right)
\label{albebound}
\end{eqnarray}
where $C(\rho)$ denotes the concurrence of a mixed bipartite quantum state $\rho$ and other notations were defined earlier.\\
We have modified the above lower bound of concurrence and obtained the modified lower bound by using the constructed witness operator.
We take few steps forward in this direction of research by obtaining a new improved lower bound of concurrence using the constructed witness operator.\\
The work is organised as follows: In section-II, we have discussed few theorems and results that have already been established in the literature. Then with the help of those results, we have derived few new results in this section. In section-III, we have constructed witness operators that can detect NPTES and PPTES. In section-IV, we have stated and proved a theorem which provides a new lower bound of the concurrence based on the constructed witness operator in the previous section. In section-V, we have discussed few examples of NPTES and PPTES in which we have shown that the constructed witness operator not only detect those entangled states but also it helps in achieving the better value of the lower bound of the concurrence of the given NPTES and PPTES. In section-VI, we end with concluding remarks.
\section{Preliminaries}
In this section, we have stated theorems and results which are discussed in the literature. Then using these results and theorems, we derive a few new theorems and results which will be used in the later section.\\

\subsection{Previous Theorems and Results}

\textbf{Theorem-P1 \cite{zou}:} Let $M_{n}(\mathbb{C})$ be the set of all complex matrices of order $n$ and $M \in M_{n}(\mathbb{C})$, then
\begin{eqnarray}
|Tr(M)|^2 \leq rank (M)  \sum_{i=1}^s |\lambda_i|^2
\label{thm1}
\end{eqnarray}
where $\lambda_i$ denotes the $i^{th}$ eigenvalue of the matrix $M$ and $s$ denotes the number of non-zero eigenvalues.

\bigskip

\textbf{Result-P1 \cite{zhan}:} Let $M_{n}$ be the set of all $n\times n$ matrices. For any $A \in M_{n}$, we have
\begin{eqnarray}
|Tr(A)| \leq \sum_{i=1}^{n} \Lambda_{i}(A)=\|A\|_{1}
\label{resp1}
\end{eqnarray}
where $\Lambda_i(A)$ denotes the $i^{th}$ singular value of the matrix $A$.
\bigskip

\textbf{Result-P2 \cite{zou}:} Let $A \in M_{n}(\mathbb{C})$, and $\lambda_{j}=a_{j}+ \iota b_{j} (j=1,2,......n)$ be an eigenvalue of $A$. Then
\begin{eqnarray}
\sum_{i=1}^{n}|\lambda_{i}|^{2} \leq \|A\|_{2}^{2}
\label{resp2}
\end{eqnarray}
where $\|.\|_{2}$ denotes the Frobenius norm defined as $\|H\|_{2}^{2}=Tr(HH^{\dagger})$.

\bigskip

\textbf{Result-P3 \cite{lin}:} If a state described by the density operator $\rho$ in $d_{1}\otimes d_{2}$ dimensional system represents a positive partial transpose (PPT) state then the following inequalities hold
\begin{eqnarray}
det(I_{d_{2}}+Tr_{A}(\rho))\leq det(I_{d_{1}d_{2}}+\rho)
\label{resp3}
\end{eqnarray}
\begin{eqnarray}
det(I_{d_{1}}+Tr_{B}(\rho))\leq det(I_{d_{1}d_{2}}+\rho)
\label{resp3i}
\end{eqnarray}
where $Tr_{A}(\rho)$ and $Tr_{B}(\rho)$ represent the partial traces of the state $\rho$ with respect to the subsystems $A$ and $B$, respectively, while $I_d$ denotes the $d\times d$ identity matrix and $det$ denotes the matrix determinant operation.
\bigskip

\textbf{Result-P4 \cite{mintert2}:}
If $W$ represents the witness operator that detects the entangled quantum state described by the density operator $\rho$ and
$C(\rho)$ denotes the concurrence of the state $\rho$ then the lower bound of concurrence is given by
\begin{eqnarray}
C(\rho)\geq -Tr[W\rho]
\label{resultP4}
\end{eqnarray}

\subsection{New Theorems and Results}

\textbf{Result-1:} If $\rho^{R}$ denotes the realigned matrix of the state described by the density operator $\rho$ in $d_{1}\otimes d_{2}$ dimensional system, then,
\begin{equation}
\|\rho^{R}\|_{1} \leq \sqrt{rank(\rho^{R})}\;  \|\rho^{R}\|_{2}
\label{res2}
\end{equation}
where $\|.\|_1$ denotes the trace norm and $\|.\|_2$ denotes the Frobenius norm.\\
\textbf{Proof:} Since $\rho^{R}$ describe the realigned matrix of the density operator $\rho$ so $Tr(\rho^{R})$ may or may not be equal to unity. Thus, replacing the complex matrix $M$ by $\rho^{R}$ in (\ref{thm1}), we have
\begin{eqnarray}
|Tr(\rho^R)|^2 \leq rank(\rho^R)  \sum_{i=1}^s |\lambda_i(\rho^R)|^2
\label{eqn1}
\end{eqnarray}
If we assume that $\rho^R$ is hermitian with positive eigenvalues, then
\begin{eqnarray}
Tr(\rho^R) = \sum_{i} \lambda_i = \sum_{i} \Lambda_i = \|\rho^R\|_1
\label{eqn11}
\end{eqnarray}
where $\lambda_{i}$ and $\Lambda_{i}$ denoting the eigenvalues and singular values of $\rho^{R}$ respectively.\\
Using (\ref{eqn1}) and (\ref{eqn11}), we get
\begin{eqnarray*}
\|\rho^{R}\|_{1}^2 &\leq& rank(\rho^{R})  \sum_{i=1}^s |\lambda_i(\rho^R)|^2 \\
&\leq& rank(\rho^{R})\;  \|\rho^{R}\|_{2}^2
\end{eqnarray*}
The last inequality follows from (\ref{resp2}). Hence proved.\\

Let us define an operator $A$ of the form
\begin{equation}
A = \frac{1}{\sqrt{rank(\rho^{R})}\;  \|\rho^{R}\|_{2}} \; \rho^{R}
\end{equation}
Using (\ref{res2}), it can be shown that $\|A\|_{1}\leq 1$.

\medskip

\textbf{Theorem-1:} If the bipartite state described by the density operator $\rho$ in $d_{1} \otimes d_{2}$ dimensional system, is separable then
\begin{equation}
\|A\|_{1} \leq \frac{1}{\sqrt{rank(\rho^{R})}\;  \|\rho^{R}\|_{2}}
\label{res3}
\end{equation}
\textbf{Proof:} Using the fact that if the state $\rho$ is separable then $\|\rho^{R}\|_{1}\leq 1$, one can prove that the $Theorem-1$ is indeed true.

\medskip

\textbf{Corollary-1:} If the inequality (\ref{res3}) is violated by a quantum state $\rho$ then the state $\rho$ must be entangled, i.e., if the
state $\rho$ satisfies
\begin{equation}
\frac{1}{\sqrt{rank(\rho^{R})}\;  \|\rho^{R}\|_{2}} < \|A\|_{1}
\label{res31}
\end{equation}
then the state $\rho$ is entangled.

\medskip

\textbf{Theorem-2:} Let $\rho^{R}$ be the realigned matrix of the bipartite state described by the density operator $\rho$ in $d_{1} \otimes d_{2}$ dimensional system. If the state $\rho$ is separable then,
\begin{equation}
{\| \rho^{T_B} \rho^{R} \|}_{1} \leq \Lambda_{max}(\rho^{T_B}) \label{res4}
\end{equation}
where $T_{B}$ denotes the partial transposition with respect to the system $B$ and $\Lambda_{max}(\rho^{T_B})$ denotes the maximum singular value of $\rho^{T_B}$.\\
\textbf{Proof:} Let us consider the product of two matrices $\rho^{T_{B}}$ and $\rho^{R}$ and further suppose that $\Lambda_i(\rho^{T_B} \rho^{R})$ denoting the $i^{th}$ singular value of the product $\rho^{T_B} \rho^{R}$. Therefore, the upper bound of trace norm of $\rho^{T_B} \rho^{R}$ is given by
\begin{eqnarray}
\| \rho^{T_B} \rho^{R} \|_1 &=& \sum \Lambda_i (\rho^{T_B} \rho^{R})
\nonumber\\&\leq& \sum \Lambda_i (\rho^{T_B})\: \Lambda_i (\rho^{R})
\nonumber\\&\leq& \Lambda_{max}(\rho^{T_B}) \sum \Lambda_i (\rho^{R})
\nonumber\\&=& \Lambda_{max}(\rho^{T_B}) \| \rho^{R} \|_{1}
\label{productub}
\end{eqnarray}
where the first inequality follows from \cite{horn}.\\
If the state $\rho$ is separable, then $ \|\rho^{R}\|_{1} \leq 1$.
Thus, the inequality (\ref{productub}) for the separable state reduces to
\begin{eqnarray}
\| \rho^{T_B} \rho^{R} \|_1 \leq \Lambda_{max}(\rho^{T_B})
\label{productsepcond}
\end{eqnarray}
Hence proved.

\bigskip

\textbf{Corollary-2:} If the state $\rho$ is separable then,
\begin{equation}
 |Tr[(\rho^{R})^{T_B} \rho]|  \leq \Lambda_{max}(\rho^{T_B}) \label{cor2}
\end{equation}
\textbf{Proof:} Let us start with $|Tr[(\rho^{R})^{T_B} \rho]|$. It is given by
\begin{eqnarray*}
 |Tr[(\rho^{R})^{T_B} \rho]| = |Tr[\rho^{R} \rho^{T_B}]| &=& |Tr[\rho^{T_B} \rho^R]|  \\
&\leq& {\| \rho^{T_B} \rho^{R} \|}_{1} \leq \Lambda_{max}(\rho^{T_B})
\end{eqnarray*}
The first inequality follows from (\ref{resp1}) and the last inequality follows from $Theorem-2$.

\section{Construction of Witness Operator}
In this section, we will construct different types of witness operators that can detect (i) only NPTES (ii) Both NPTES and PPTES.

\subsection{Witness operator detecting only NPTES}
\noindent It is well known that partial transposition operation can detect NPTES but the problem lies in the fact that it is not a physical operation and thus not possible to implement it in real experiment. To resolve this issue, we take an approach of constructing a witness operator that does not contain partial transposition map for the detection of NPTES.\\
Let us consider a $d_{1}\otimes d_{2}$ dimensional quantum state described by the density operator $\rho$. Our task is to determine whether the state described by the density operator $\rho$ is NPTES.
\medskip

\textbf{Theorem-3:} A $d_{1}\otimes d_{2}$ dimensional quantum state $\rho$ is NPTES if there exist a witness operator $\tilde{W}$ such that
\begin{eqnarray}
Tr(\tilde{W}\rho)<0
\label{theorem1}
\end{eqnarray}
where $\tilde{W}$ is given by
\begin{eqnarray}
\tilde{W}&=& \frac{det(I_{d_{1}d_{2}}+(.))}{\lambda}|\psi\rangle\langle\psi|\nonumber\\&& - (det(I_{d_{2}}+Tr_{A}(.)))I_{d_{1}d_{2}}
\label{witnptes1}
\end{eqnarray}
$(.)$ means a $d_{1}\otimes d_{2}$ dimensional bipartite state which is under investigation, $Tr_{A}(.)$ represents the partial trace with respect to the subsystem $A$ of the state under investigation, $I_{d_{1}d_{2}}$ denoting the identity matrix in $d_{1}\otimes d_{2}$ dimensional Hilbert space and $|\psi\rangle$ be the normalized eigenvector corresponding to any non-zero eigenvalue $\lambda$ of $\rho$.\\
\textbf{Proof:} Let us consider any separable state $\sigma$ in $d_{1}\otimes d_{2}$ dimensional system. The trace of the operator $\tilde{W}$ over a separable state $\sigma$ is given by
\begin{eqnarray}
Tr(\tilde{W}\sigma)&=& Tr[(\frac{det(I_{d_{1}d_{2}}+\sigma)}{\lambda}|\psi\rangle\langle\psi| \nonumber\\&&- det(I_{d_{2}}+Tr_{A}(\sigma))I_{d_{1}d_{2}})\sigma]\nonumber\\&=& det(I_{d_{1}d_{2}}+\sigma)-det(I_{d_{2}}+Tr_{A}(\sigma))
\nonumber\\&\geq& 0
\label{cond1}
\end{eqnarray}
The last step follows from (\ref{resp3}). Since $\sigma$ is an arbitrary separable state so $Tr(\tilde{W}\sigma)\geq 0$ for
any separable state.\\
Next, let us consider a state described by the density operator $\rho_{12}$ defined as
\begin{eqnarray}
\rho_{12}=
\begin{pmatrix}
  \frac{13}{30} & 0 & 0 & \frac{11}{30} \\
  0 & \frac{1}{15} & 0 & 0 \\
  0 & 0 & \frac{1}{15} & 0 \\
  \frac{11}{30} & 0 & 0 & \frac{13}{30}
\end{pmatrix}
\label{state1}
\end{eqnarray}
It can be easily shown that the state $\rho_{12}$ is indeed an entangled state.\\
The trace value of $\tilde{W}$ with respect to the state $\rho_{12}$ is given by
\begin{eqnarray}
Tr(\tilde{W}\rho_{12}) = -\frac{491}{7500}<0
\label{trdet}
\end{eqnarray}
Thus, the operator $\tilde{W}$ is a witness operator.

\bigskip

We note that if $\rho$ denotes the PPT entangled state in $d_1 \otimes d_2$ ($d_1, d_2 \geq 3$) dimensional system, then it can be easily shown that $Tr(\tilde{W}\rho)\geq 0$. This happens because (\ref{resp3}) holds for any PPTES also. Thus, it is not possible to detect any PPTES using the witness operator $\tilde{W}$. Therefore, the witness operator $\tilde{W}$ detect only NPTES.

\medskip

Let us now consider a family of $3\otimes 3$ dimensional isotropic state \cite{zhao}, which is defined by
\begin{eqnarray}
\rho_{iso}(f)=\frac{1-f}{8}I_{9}+\frac{9f-1}{8}|\psi^{+}\rangle\langle\psi^{+}|, 0\leq f\leq 1
\label{isotropic}
\end{eqnarray}
where $|\psi^{+}\rangle=\frac{1}{\sqrt{3}}(|00\rangle+|11\rangle+|22\rangle)$ and $f=\langle\psi^{+}|\rho_{iso}(f)|\psi^{+}\rangle$.\\
The state $\rho_{iso}(f)$ is separable when $f\leq\frac{1}{3}$ and NPTES when $f>\frac{1}{3}$.\\
We now calculate $Tr(\tilde{W}\rho_{iso}(f))$ to determine how efficiently $\tilde{W}$ detect NPTES.
$Tr(\tilde{W}\rho_{iso}(f))$ is given by
\begin{eqnarray}
Tr(\tilde{W}\rho_{iso}(f))&=& \frac{(-9+f)^{8}(1+f)}{16777216}-\frac{64}{27}
\nonumber\\&<&0,~~~~~0.591634<f\leq 1
\label{traceisotropicdet}
\end{eqnarray}
Thus, the witness operator $\tilde{W}$ fails to detect a few members in the family of isotropic NPTES. When the parameter $f$ lies in the region $\frac{1}{3}<f\leq 0.591634$, the corresponding entangled states are not detected by $\tilde{W}$. Hence we can say that the witness operator $\tilde{W}$ is not as much as efficient in comparison to the other witnesses in the literature.\\

Here one may argue about the utility of constructing $\tilde{W}$ to detect NPTES for which we already have the partial transposition (PT) criterion. It is known that the partial transposition is not a completely positive map and thus it would be very difficult to implement it in the laboratory. One approach to overcome this complication was given in \cite{anu} where the method of structural physical approximation of a partial transposition (SPA-PT) is adopted to detect NPTES. We provide a complementary approach to address this problem by constructing the witness operator $\tilde{W}$, which is independent of the PT operation. Our criterion  is constructive and applicable to detect NPTES even in the higher dimensional bipartite systems where the SPA method can be strenuous to implement.\\

Now, our task is to construct another witness operator that can be as efficient as $\tilde{W}$. To achieve this, let us start with $\rho^R$, which  denotes the realigned matrix of the state under investigation. Then the operator $W^{o}$ can be defined as
\begin{eqnarray}
W^{o} = \left(1 + \frac{1 - \|\rho^{R}\|_{1}}{\sqrt{rank(\rho^{R})}\;  \|\rho^{R}\|_{2}}\right)I_{d^{2}} -\frac{(\rho^{R})^{T_B}}{\Lambda_{max}(\rho^{T_B})}
\label{wit2}
\end{eqnarray}
where $T_{B}$ is the partial transpose with respect to the second subsystem $B$ and $\Lambda_{max}(\rho^{T_B})$ denotes the maximum singular value
of $\rho^{T_B}$.\\
\textbf{Theorem-4:} The operator $W^o$ is an entanglement witness operator.\\
\textbf{Proof:} Let us consider any $d_{1}\otimes d_{2}$ dimensional bipartite separable state $\sigma$. Therefore, $Tr(W^{o} \sigma)$ is given by
\begin{eqnarray}
Tr(W^{o} \sigma)= 1 - \frac{Tr[\sigma^{T_B} \sigma^R]}{\Lambda_{max}(\sigma^{T_B})}  +  \frac{1 - \|\sigma^{R}\|_{1}}{\sqrt{rank(\sigma^{R})}\|\sigma^{R}\|_2}
\label{tracevalueWo}
\end{eqnarray}
From (\ref{res3}) and (\ref{cor2}) it follows that $Tr(W^{o}\sigma)\geq 0$ for all separable states.\\
Now it remains to show that there exists at least one entangled state $\rho$ for which $Tr(W^{o} \rho)< 0$. For this, let us
consider a state of the form
\begin{eqnarray}
\varrho_{12}=
\begin{pmatrix}
  \frac{11}{30} & 0 & 0 & \frac{7}{30} \\
  0 & \frac{2}{15} & 0 & 0 \\
  0 & 0 & \frac{2}{15} & 0 \\
  \frac{7}{30} & 0 & 0 & \frac{11}{30}
\end{pmatrix}
\label{state2}
\end{eqnarray}
It can be easily verified that the state $\varrho_{12}$ is an entangled state.\\
The quantity $Tr(W^{o}\varrho_{12})$ is given by
\begin{eqnarray}
Tr(W^{o}\varrho_{12})= -0.0585731 < 0
\label{tr3}
\end{eqnarray}
Thus, the operator $W^{o}$ is indeed an entanglement witness operator.\\
Let us now recall again the family of $3\otimes3$ isotropic states defined in (\ref{isotropic}) and investigate whether
the witness operator $W^{o}$ detect more members of the family of isotropic states than $\tilde{W}$. To probe this, let us calculate the following: \\
\begin{eqnarray*}
Tr[({\rho^{T_B}_{iso}} \; \rho^R_{iso})(f)] &=& \frac{1}{96} ( -1 + 42f - 9f^2)\\
\Lambda_{max}(\rho^{T_B}_{iso}(f)) &=&
\begin{cases}
\frac{1 - 3f}{6} &  0 \leq f \leq \frac{1}{9} \\
\frac{1 + 3f}{12} &  \frac{1}{9} \leq f \leq 1 \\
\end{cases}\\
rank(\rho^R_{iso}(f))&=&
\begin{cases}
1 &  f = \frac{1}{9}\\
9 & f \neq \frac{1}{9} \\
\end{cases}\\
\|\rho^R_{iso}(f)\|_{1}&=&
\begin{cases}
\frac{2}{3} - 3f &  0 \leq f \leq \frac{1}{9} \\
3f &  \frac{1}{9} \leq f \leq 1 \\
\end{cases}\\
\|\rho^R_{iso}(f)\|_{2}&=& \sqrt{\frac{1 - 2f + 9f^2}{8}}\\
\end{eqnarray*}

Using (\ref{tracevalueWo}), we get
\begin{eqnarray}
& &Tr(W^{o}\rho_{iso}(f))\nonumber\\&=&
\begin{cases}
\frac{17 - 90f + 9f^2}{16 - 48f} + \frac{2\sqrt{2}(1 + 9f)}{9\sqrt{1 - 2f + 9f^2}}& 0 \leq f < \frac{1}{9}\\
\frac{8}{3} & f = \frac{1}{9}\\
\frac{1}{3}\left(\frac{27 (-1 + f)^2}{8 + 24f} - \frac{2\sqrt{2} (-1 + 3f)}{\sqrt{1 - 2f + 9f^2}}\right) & \frac{1}{9} < f \leq 1
\end{cases}
\label{traceWo_iso}
\end{eqnarray}
Here, $Tr(W^{o}\rho_{iso}(f)) < 0$ for $0.413285 < f \leq 1$, which improves the detection range obtained in (\ref{traceisotropicdet}). Thus, the witness operator $W^{o}$ can be considered as more efficient than the witness operator $\tilde{W}$. We can now observe the following facts:\\
(i) $W^{o}$ may detect bound entangled states also.\\
(ii) $\rho^R$ and $({\rho^R})^{T_B}$ are both non-Hermitian matrices. But the real eigenvalues of $({\rho^R})^{T_B}$ makes our witness operator CPT symmetric and capable of detecting entanglement \cite{bender, bender05, pati}. In most of the cases, we find that the eigenvalues of $({\rho^R})^{T_B}$ are real.

\subsection{Witness operator detecting both NPTES and PPTES}
In this section, our task is to construct a witness operator which is efficient in detecting both NPTES and PPTES.\\
Let us now start with the operator $W_{(n)}$ which can be defined as follows:
\begin{eqnarray}
W_{(n)} = \frac{d_{1}}{d_{1}-1} \left[
\begin{multlined}
(k_{\rho})^{n} \left(I_{d_{1}d_{2}} - \frac{(\rho^{R})^{T_B}}{\Lambda_{max}(\rho^{T_B})}\right) \\+ \left(\frac{1 - \|\rho^{R}\|_{1}}{\sqrt{rank(\rho^{R})}\;  \|\rho^{R}\|_{2}}\right)I_{d_{1}d_{2}}
\end{multlined}
\right] \label{witn1}
\end{eqnarray}
where $n \in \mathbb{N}$, the set of natural numbers; and\\ $k_{\rho} = det(I_{d_{1}d_{2}}+\rho) - det(I_{d_{2}} + Tr_{A}(\rho))$.\\
\textbf{Theorem-5:} The operator $W_{(n)}$ is a witness operator that can detect PPTES.\\
\textbf{Proof:} Let us consider a bipartite $d_{1} \otimes d_{2}$ $(d_{1}\leq d_{2})$ dimensional separable state $\sigma$. Using (\ref{resp3}), we can have $k_{\sigma} \geq 0$ for any separable state $\sigma$.
\begin{eqnarray}
Tr(W_{(n)} \sigma)= \frac{d_{1}}{d_{1}-1}\left[
\begin{multlined}
(k_{\sigma})^{n}\left(1 - \frac{Tr[\sigma^{T_B} \sigma^R]}{\Lambda_{max}(\sigma^{T_B})}\right) \\+  \frac{1 - \|\sigma^{R}\|_{1}}{\sqrt{rank(\sigma^{R})}\|\sigma^{R}\|_2}
\end{multlined}
\right]
\label{tracevalueWn}
\end{eqnarray}
Using (\ref{tracevalueWo}), it follows that $Tr(W_{(n)}\sigma)\geq 0$ for all bipartite $d_{1} \otimes d_{2}$ dimensional separable state $\sigma$.\\
Let us now consider the bound entangled state given in \cite{bihalan}
\begin{equation}
\rho_{BE} =
\begin{pmatrix}
a&0&0&0&b&0&0&0&b\\
0&c&0&0&0&0&0&0&0\\
0&0&a&0&0&0&0&0&0\\
0&0&0&a&0&0&0&0&0\\
b&0&0&0&a&0&0&0&0\\
0&0&0&0&0&c&0&b&0\\
0&0&0&0&0&0&c&0&0\\
0&0&0&0&0&b&0&a&0\\
b&0&0&0&0&0&0&0&a\\
\end{pmatrix}
\end{equation}
where $a=\frac{1+\sqrt{5}}{3+9\sqrt5}$, $b=\frac{-2}{3+9\sqrt5}$, $c=\frac{-1+\sqrt{5}}{3+9\sqrt5}$. The values of the parameters involved in the witness operator $W_{(n)}$ to detect the state $\rho_{BE}$ are given below.\\
\begin{eqnarray}
&&d_{\rho_{BE}}=0.149 > 0, Tr[\rho_{BE}^{T_B} \rho_{BE}^R] = \frac{1}{363} (21 - 8\sqrt{5}), \nonumber\\&&
\Lambda_{max} (\rho_{BE}^{T_B})= \frac{1}{33} \sqrt{29 + 12\sqrt{5}}, rank(\rho_{BE}^R) = 9, \nonumber\\&&
\|\rho_{BE}^R\|_1 = 1.025, \|\rho_{BE}^R\|_2 = 0.413
\end{eqnarray}
The expectation value of $W_{(n)}$ with respect to the state $\rho_{BE}$ is given by
\begin{eqnarray}
Tr(W_{(n)} \rho_{BE}) &=& 1.5 (0.0203459 - 0.962145 \times 0.149599^n) \nonumber\\ &<&  0~~
\text{for}~~ \; n \geq 3
\end{eqnarray}
Since the operator $W_{(n)}$ detects the PPTES described by the density operator $\rho_{BE}$ for each $n \geq 3$ so $W_{(n)}$ is a witness
operator. Hence proved.

\section{Lower Bound of the concurrence}
In this section, we have derived a new lower bound of concurrence of a bipartite quantum state $\rho$ in $d_{1} \otimes d_{2}$ dimensional system and show that our bound is better in most cases when it is compared to the lower bound of the concurrence given by \cite{kchen}. We note that the lower bound given in (\ref{albebound}) is not normalized but can be normalized to unity. If $C_{min}(\rho)$ denotes the normalized value of this bound for the state $\rho$, then we have
\begin{eqnarray}
C(\rho)&\geq& C_{min}(\rho)\nonumber\\ &=& \frac{1}{(d_{1}-1)}(max(\|\rho^{T_{A}}\|_{1},\|\rho^{R}\|_{1})-1)
\label{nalbebound12}
\end{eqnarray}
We are now in a position to use the witness operator $W_{(n)}$ defined in (\ref{witn1}) in the result-P4 by Mintert \cite{mintert2} for getting the improvement of the lower bound of the concurrence of an arbitrary bipartite $d_{1}\otimes d_{2}$ dimensional system. It may be noted that not all witness operators improve the lower bound of the concurrence given by (\ref{albebound}). 
\medskip

\textbf{Theorem-6:} Let $\rho$ be an entangled state in $d_{1}\otimes d_{2}$ $(d_{1}\leq d_{2})$ dimensional system detected by the witness operator $W_{(n)}$ defined in (\ref{witn1}). Then there exist $n_{1}\in \mathbb{N}$ such that the lower bound of concurrence of the state $\rho$ is given by
\begin{eqnarray}
C(\rho)\geq \Phi_{W_{(n)}} (\rho),~~ \forall n\geq n_{1}
\label{result6}
\end{eqnarray}
where
\begin{eqnarray}
\Phi_{W_{(n)}} (\rho) = - Tr[W_{(n)} \rho] &=&
\frac{d_{1}}{d_{1}-1} [(k_{\rho})^{n}(\frac{Tr[\rho^{T_B} \rho^R]}{\Lambda_{max}(\rho^{T_B})} - 1) \nonumber\\&+& \frac{\|\rho^{R}\|_{1} - 1}{\sqrt{rank(\rho^{R})} \|\rho^{R}\|_2}]
\label{ourbound}
\end{eqnarray}
\textbf{Proof:} Let us first recall the witness operator $W_{(n)}$ defined in (\ref{witn1}). Then the theorem follows by using the witness operator $W_{(n)}$ in the result given in (\ref{resultP4}). Hence proved.\\
\textbf{Lemma-1:} For any bipartite state $\rho$ in $d_1 \otimes d_2$ dimensional system, we have
\begin{eqnarray}
|k_{\rho}| < 1
\end{eqnarray}
where $k_{\rho} = det(I_{d_{1}d_{2}}+\rho) - det(I_{d_{2}} + Tr_{A}(\rho))$.\\
\textbf{Proof:} Let us start with the expression of $Tr(I_{d_1 d_2} + \rho)$. The value of $Tr(I_{d_1 d_2} + \rho)$ is given by
\begin{eqnarray}
Tr(I_{d_1 d_2} + \rho) = Tr(I_{d_1 d_2}) + Tr(\rho) = d_1d_2+1
\label{l1}
\end{eqnarray}
Moreover, the inequality $det(I_{d_{1}d_{2}}+\rho) \leq \left(\frac{1 + d_1d_2}{d_1d_2}\right)^{d_1d_2}$ can be derived as
\begin{eqnarray}
1 + d_1d_2 &=& Tr(I_{d_1 d_2} + \rho) \nonumber\\
&=&  \sum_{i=1}^{d_1d_2} \lambda_i(I_{d_1 d_2} + \rho) \nonumber\\
&\geq&  d_1d_2 \left(\prod_{i=1}^{d_1d_2} \lambda_i(I_{d_1 d_2} + \rho)\right)^{\frac{1}{d_1d_2}}  \nonumber\\
&=&  d_1d_2 \left(det(I_{d_{1}d_{2}}+\rho)\right)^{\frac{1}{d_1d_2}}
\end{eqnarray}
i.e.,
\begin{equation}
det(I_{d_{1}d_{2}}+\rho) \leq \left(\frac{1 + d_1d_2}{d_1d_2}\right)^{d_1d_2}
\label{expr1}
\end{equation}
It can be seen that R.H.S of (\ref{expr1}) tends toward \textit{Euler's} number \textit{e} as $d_1,d_2$ tends to $\infty$.\\
Therefore, for arbitrary large value of $d_{1}$ and $d_{2}$, we have
\begin{equation}
det(I_{d_{1}d_{2}}+\rho) \leq \textit{e}
\label{expr11}
\end{equation}
Let us first calculate the bound of $det(I_{d_{2}} + Tr_{A}(\rho))$ for $d_{2}=2$ and then generalize the result to arbitrary dimension $d_{2}$.
The quantum state in 2-dimensional system, i.e., a qubit is described by the density operator
\begin{equation}
\varrho^{(2)} = \frac{I_2 + \vec{r}.\vec{\sigma}}{2}
\label{expr11}
\end{equation}
where $\vec{r} \in \mathbb{R}^3$ with $|\vec{r}|^2 \leq 1$ is the \textit{Bloch vector} for the state $\varrho^{(2)}$; $I_2$ is $2\times2$ identity matrix; and $\vec{\sigma} = (\sigma_x,\sigma_y,\sigma_z)$ where $\sigma_x,\sigma_y$ and $\sigma_z$ are \textit{Pauli matrices} \cite{chuang}.\\
After carrying out a simple calculations, we arrive at the result given by
\begin{equation}
det(I_2 + \varrho^{(2)}) \geq 2
\label{expr12}
\end{equation}
The equality holds in (\ref{expr12}) for pure states.\\
Since pure states are rank one projectors, so we have $det(I_{d_{2}} + \varrho^{(d_2)}) = 2$ for any pure state $\varrho^{(d_2)}$ in $d_2$ dimensional system. Thus, we can generalize the result (\ref{expr12}) to an arbitrary qudit described by the density operator $\varrho^{(d_2)}$, we obtain the following
\begin{equation}
det(I_{d_{2}} + \varrho^{(d_2)}) \geq 2 \label{quditcase}
\end{equation}
Using the results (\ref{expr1}) and (\ref{quditcase}) in $k_{\rho}$, we get
\begin{eqnarray}
k_{\rho} &=& det(I_{d_{1}d_{2}}+\rho) - det(I_{d_{2}} + Tr_{A}(\rho)) \nonumber\\
&\leq& \textit{e} - 2 < 1
\end{eqnarray}
Similarly, we can show that $ k_{\rho} > -1$. Thus, we have $|k_{\rho}|<1$. Hence proved.\\

 \textbf{Note-1:} For PPT states, we have $0 \leq k_{\rho} < 1$.\\

\noindent \textbf{Corollary-3:} For large value of $n$ i.e. as $n\rightarrow \infty$, the lower bound of concurrence is given by
\begin{eqnarray}
C(\rho)\geq \phi(\rho)
\label{result7}
\end{eqnarray}
where $\quad \phi(\rho) = \frac{d_{1}}{d_{1}-1}  \left(\frac{\|\rho^{R}\|_{1} - 1}{\sqrt{rank(\rho^{R})} \|\rho^{R}\|_2}\right)$.\\
\textbf{Proof:} Since $k_{\rho}<1$ so $(k_{\rho})^{n}\rightarrow 0$, as $n\rightarrow \infty$. Thus, we have
\begin{eqnarray}
\lim_{n\rightarrow\infty}\Phi_{W_{(n)}} (\rho) = \phi(\rho) \label{limvalue}
\end{eqnarray} \\
Hence proved.\\

\textbf{Note-2:} The lower bound of concurrence given in Eq. (\ref{ourbound}) is better than that given in Eq. (\ref{nalbebound12}), when $n\rightarrow \infty$.

\section{Examples}
 In this section, we will discuss some examples to illustrate the utility of the witness operator $W_{(n)}$ in detecting NPTES and PPTES. Furthermore, we will improve the lower bound of the concurrence of given NPTES and PPTES detected by the witness operator $W_{(n)}$. Also, we have shown that for the given state $\rho$ the lower bound of the concurrence $\Phi_{W_{(n)}} (\rho)$ defined in (\ref{ourbound}) tends towards $\phi(\rho)$ for sufficiently large value of $n$. In this context, we have provided few examples in which the following relation holds:
\begin{eqnarray}
C(\rho)\geq \Phi_{W_{(n)}} (\rho)\geq C_{min}(\rho) ,~~ \forall n\geq n_{1}
\label{result60}
\end{eqnarray}
Using $Corollary-3$, we have the following relation:
\begin{eqnarray}
C(\rho)\geq \phi(\rho)\geq C_{min}(\rho)
\label{result60}
\end{eqnarray}
\subsection{Detection and Estimation of Lower Bound of Concurrence of NPTES}
\textbf{Example-1:} Let us again recall the $3\otimes3$ isotropic states described by the density operator $\rho_{iso}(f)$ defined in (\ref{isotropic}). The witness operator $W_{(n)}$ detects $3\otimes3$ isotropic states for some range of the parameters which has been shown in Table-\ref{table_iso} given below:
\begin{table}[h!]
\begin{tabular}{| p{1cm} | p{3cm} | p{3cm} |}
\hline
\multicolumn{3}{|c|}{$3\otimes 3$ Isotropic States described by $\rho_{iso}(f)$} \\
\hline
\textit{n} & The range of the parameter $f$ for which $Tr(W_{(n)} \rho_{iso}(f))<0$ &$\rho_{iso}(f)$ detected by the witness operator $W_{(n)}$ \\
\hline
1 &  $0.35 < f \leq 1$  &  NPTES detected by $W_{(1)}$ \\
\hline
2 &  $0.336 < f \leq 1$  &  NPTES detected by $W_{(2)}$\\
\hline
3 &  $0.3338 < f \leq 1$  &  NPTES detected by $W_{(3)}$\\
\hline
4 &  $0.3334 < f \leq 1$  &  NPTES detected by $W_{(4)}$\\
\hline
5 &  $0.33334 < f \leq 1$  &  NPTES detected by $W_{(5)}$\\
\hline
\end{tabular}
\caption{Detection of isotropic state using $W_{(n)}$ in the range \\$\frac{1}{3} < f \leq 1$}
\label{table_iso}
\end{table} \\
Table-\ref{table_iso} shows that the range of the parameter $f$ to detect $3\otimes3$ negative partial transpose isotropic states (\ref{isotropic}) increases as $n$ increases.\\
We will now use the witness operator $W_{(n)}$ to estimate the lower bound of the concurrence of $\rho_{iso}(f)$. With an increase in $n$, one can easily find the improvement in the lower bound of concurrence estimated by the witness operator $W_{(n)}$  $\forall \; n \in \mathbb{N}$, when compared to the lower bound of the concurrence given in (\ref{nalbebound12}). If we take sufficiently large value of $n$ then from $Corollary-3$, we have
\begin{eqnarray}
C(\rho_{iso}(f)) \geq \phi(\rho_{iso}(f)) &=& \frac{\sqrt{2} (-1 + 3f)}{\sqrt{1 - 2f + 9f^2}},
\nonumber\\&& \frac{1}{3}<f\leq 1
\label{lbcex-1}
\end{eqnarray}
where $C(\rho_{iso}(f))$ denotes the concurrence of the isotropic state.\\
In Figure-1, we have compared the lower bound $\phi(\rho_{iso})$ given in (\ref{lbcex-1}) with the lower bound $C_{min}(\rho_{iso})$ given in (\ref{nalbebound12}).
\begin{figure}[h!]
\includegraphics[width=0.45\textwidth]{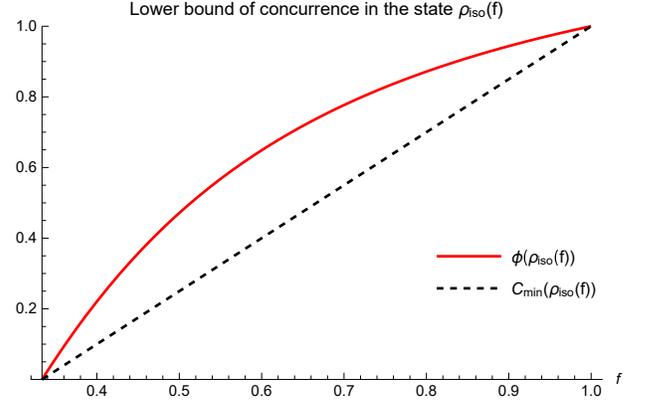}
\caption{Isotropic states: The dotted curve represents $C_{min}(\rho_{iso})$ and the solid red curve represents the limiting value of our bound, i.e., $\phi(\rho_{iso})$. Clearly, in the entangled region $\frac{1}{3}< f \leq 1$, $\phi(\rho_{iso})$ gives a better estimate of the lower bound of concurrence as compared to $C_{min}(\rho_{iso})$.} \label{fig1}
\end{figure}
\\

\begin{figure}[h!]
\includegraphics[width=0.45\textwidth]{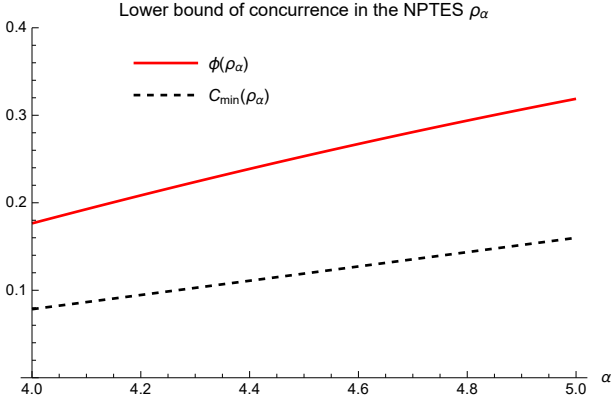}
\caption{Horodecki alpha states: The dotted curve represents $C_{min}(\rho_{\alpha})$ and the solid red curve represents the limiting value of our bound, i.e., $\phi(\rho_{\alpha})$. Clearly, in the NPT entangled region $4 < \alpha \leq 5$, $\phi(\rho_{\alpha})$ gives a better estimate of the lower bound of concurrence as compared to $C_{min}(\rho_{\alpha})$.} \label{fig2}
\end{figure}

\textbf{Example-2:} Let us consider a class of bipartite quantum state in $3\otimes 3$ dimensional system, which is defined as \cite{horodecki1}
\begin{eqnarray}
\rho_{\alpha}=\frac{2}{7}|\psi^{+}\rangle\langle\psi^{+}|+\frac{\alpha}{7}\sigma_{+}+\frac{5-\alpha}{7}\sigma_{-},~~~2\leq \alpha \leq 5
\label{pptes2}
\end{eqnarray}
where $|\psi^{+}\rangle=\frac{1}{\sqrt{3}}(|00\rangle+|11\rangle+|22\rangle)$ and
\begin{eqnarray}
\sigma_{+}=\frac{1}{3}(|01\rangle\langle01|+|12\rangle\langle12|+|20\rangle\langle20|)
\label{sigma+}
\end{eqnarray}
\begin{eqnarray}
\sigma_{-}=\frac{1}{3}(|10\rangle\langle10|+|21\rangle\langle21|+|02\rangle\langle02|)
\label{sigma-}
\end{eqnarray}
The state $\rho_{\alpha}$ can be characterized with respect to the parameter $\alpha$ in the interval $[2,5]$  as:\\
(i) $\rho_{\alpha}$ is a separable state when $2\leq \alpha \leq 3$.\\
(ii) $\rho_{\alpha}$ represents PPTES when $3<\alpha \leq 4$.\\
(iii) $\rho_{\alpha}$ is NPTES when $4< \alpha \leq 5$. \\
In this example, we will consider the state $\rho_{\alpha}$ for $4< \alpha \leq 5$.\\
It can be easily seen that for each $n$, the witness operator $W_{(n)}$ detects all the NPTES belonging to the family of states described by the density operator $\rho_{\alpha},~~4< \alpha\leq 5$. Further, we can use the witness operator $W_{(n)}$ to improve the estimation of lower bound of concurrence of the state $\rho_{\alpha},~~4< \alpha\leq 5$. In this case, we find that except for $n=1$, the witness operator $W_{(n)}$ improves the lower bound of the concurrence compared to the lower bound given in (\ref{nalbebound12}) in the whole range of the parameter $\alpha$, i.e., $4< \alpha \leq 5$. For $n=1$, the witness operator $W_{(1)}$ improves the lower bound in the interval $4.15 < \alpha \leq 5$. Figure-\ref{fig2} describes a comparison between the limiting value of our bound, i.e., $\phi(\rho_{\alpha})$ with $C_{min}(\rho_{\alpha})$.
\subsection{Detection and Estimation of Lower Bound of Concurrence of PPTES}
\noindent We will now consider few examples of bound entangled states detected by the witness operator $W_{n}$.\\

\textbf{Example-1:} A $3\otimes 3$ BES constructed from the unextendible product basis (UPB) is given by \cite{cbennett}
\begin{eqnarray}
\rho_{\mathcal{B}} = \frac{1}{4}[I_{9}-\sum_{i=1}^{5}|\psi_{i}\rangle\langle\psi_{i}|]
\label{bennetstate}
\end{eqnarray}
where the states $\{|\psi_{i}\rangle\}_{i=1}^5$ form the UPB and are given by
\begin{eqnarray}
&&|\psi_{1}\rangle=\frac{1}{\sqrt{2}}|0\rangle\otimes(|0\rangle-|1\rangle)\nonumber\\&&
|\psi_{2}\rangle=\frac{1}{\sqrt{2}}(|0\rangle-|1\rangle)\otimes|2\rangle\nonumber\\&&
|\psi_{3}\rangle=\frac{1}{\sqrt{2}}|2\rangle\otimes(|1\rangle-|2\rangle)\nonumber\\&&
|\psi_{4}\rangle=\frac{1}{\sqrt{2}}(|1\rangle-|2\rangle)\otimes|0\rangle\nonumber\\&&
|\psi_{5}\rangle=\frac{1}{3}(|0\rangle+|1\rangle+|2\rangle)\otimes(|0\rangle+|1\rangle+|2\rangle)\nonumber\\&&
\label{upb}
\end{eqnarray}
To start with, let us calculate the following quantities.
\begin{eqnarray}
&&k_{\rho_{\mathcal{B}}} = \frac{71}{768}, Tr[\rho_{\mathcal{B}}^{T_B} \rho_{\mathcal{B}}^R] = \frac{1}{16},
\Lambda_{max}(\rho_{\mathcal{B}}^{T_B}) = \frac{1}{4},\nonumber\\&& rank(\rho_{\mathcal{B}}^{R}) = 6,
\|\rho_{\mathcal{B}}^{R}\|_{1} = 1.08741, \|\rho_{\mathcal{B}}^{R}\|_{2} = 0.5
\label{i1}
\end{eqnarray}
Using the above data given in (\ref{i1}), we can construct the witness operator $W_{(n)}$ and calculate its expectation value with respect to the state $\rho_{\mathcal{B}}$ as
\begin{eqnarray}
Tr(W_{(n)} \rho_{\mathcal{B}}) &=& \frac{3}{2} \left[\begin{multlined}
(k_{\rho_{\mathcal{B}}})^{n}\left(1 - \frac{Tr[\rho_{\mathcal{B}}^{T_B} \rho_{\mathcal{B}}^R]}{\Lambda_{max}(\rho_{\mathcal{B}}^{T_B})}\right)  \nonumber\\+  \frac{1 - \|\rho_{\mathcal{B}}^{R}\|_{1}}{\sqrt{rank(\rho_{\mathcal{B}}^{R})}\|\rho_{\mathcal{B}}^{R}\|_2}\\
\end{multlined}
\right]\\
&=& \frac{3}{2} \left(\frac{3}{4}\left(\frac{71}{768}\right)^n - 0.071372\right) \nonumber\\ &<&  0 \quad \forall \; n \in \mathbb{N}
\label{tracevalueWn_bennet}
\end{eqnarray}
Thus, $W_{(n)}$ detect the BES described by the density operator $\rho_{\mathcal{B}}$ for all $n \in \mathbb{N}$.\\
The lower bound of the concurrence of the state $\rho_{\mathcal{B}}$ is given by
\begin{eqnarray}
C(\rho_{\mathcal{B}}) &\geq& \Phi_{W_{(n)}}(\rho_{\mathcal{B}}) = - Tr(W_{(n)} \rho_{\mathcal{B}}) \quad\forall \; n \in \mathbb{N}
\end{eqnarray}
In Table - \ref{Bennet_table}, we have compared the lower bound of concurrence of the state $\rho_{\mathcal{B}}$. It shows that for $n > 1$, the function $\Phi_{W_{(n)}}$ gives a better estimate of the lower bound of concurrence as compared to (\ref{nalbebound12}), i.e.,
\begin{eqnarray}
\Phi_{W_{(n)}}(\rho_{\mathcal{B}}) > C_{min}(\rho_{\mathcal{B}}) = 0.04 \quad\forall \; n > 1
\end{eqnarray}

\begin{table}[h]
\begin{tabular}{| p{1cm} | p{3cm} | p{3cm} | }
\hline
\multicolumn{3}{|c|}{Lower bound of concurrence for the state $\rho_{\mathcal{B}}$} \\
\hline
\textit{n} & $\Phi_{W_{(n)}} (\rho_{\mathcal{B}})$ & $C_{min}(\rho_{\mathcal{B}})$  \\
\hline
1 &  0.00305406 &  0.04 \\
\hline
2 &  0.097443  & 0.04 \\
\hline
3 &  0.106169  &  0.04  \\
\hline
4 &  0.106976  & 0.04 \\
\hline
5 &  0.10705  &  0.04  \\
\hline
\end{tabular}
\caption{Lower bound is compared with $C_{min}(\rho_{\mathcal{B}})$}
\label{Bennet_table}
\end{table}
Also it can be observed that as we increase the value of $n$, the value of the lower bound of concurrence is also improved. So, it would be interesting to find out the value of the lower bound of concurrence for indefinite large $n$. We calculate the lower bound of concurrence of $\rho_{\mathcal{B}}$ for large $n$ and using $Corollary-3$, it can be estimated as \\
\begin{eqnarray}
C(\rho_{\mathcal{B}})\geq \phi(\rho_{\mathcal{B}})=0.107058
\label{cor3ex1}
\end{eqnarray}\\

\begin{figure}[h!]
\includegraphics[width=0.45\textwidth]{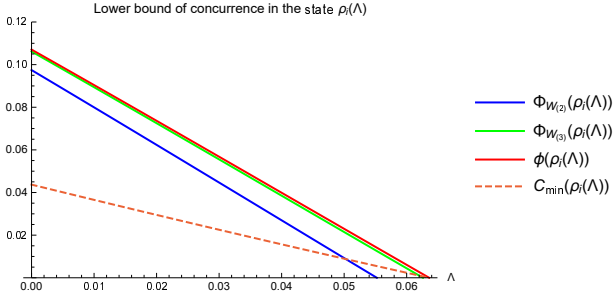}
\caption{Our lower bound $\Phi_{W_{(n)}}(\rho_i(\gamma))$, is represented by the solid curves, in blue ($n=2$), green($n=3$). One must observe that for $n>3$, the bound is slightly less than $\phi(\rho_{i}(\gamma))$ and finally converges to it, i.e., to the red curve, as n is increased. The dotted curve represents $C_{min}(\rho_i(\gamma))$, i.e., the normalized lower bound concurrence of the state $\rho_i(\gamma)$ given by (\ref{nalbebound12}).}
\label{lambdaimage}
\end{figure}

\textbf{Example-2:} Let us consider the two qutrit, non-full rank BES given by \cite{bandyo}
\begin{eqnarray}
\rho_i(\gamma) = \gamma |\psi_i\rangle \langle\psi_i| + (1-\gamma)\rho_{\mathcal{B}},~~1\leq i \leq 5
\label{ex-2b}
\end{eqnarray}
where $\rho_{\mathcal{B}}$, $|\psi_i\rangle$ are defined in (\ref{bennetstate}) and (\ref{upb}) and $\gamma \in [0,1]$.\\
The state $\rho_i(\gamma)$ satisfy the range criteria. For any $i(1\leq i \leq 5)$, the PPT states $\rho_i(\gamma)$ are bound entangled if and only if $0\leq \gamma < \frac{1}{5}$. $\rho_i(\gamma)$ represent separable states for $\frac{1}{5} \leq \gamma \leq 1 $.\\
After simple calculations, we find that our witness operator $W_{n}$ identify the states $\rho_i(\gamma),~~1\leq i \leq 5$ given in (\ref{ex-2b}) as bound entangled in the region $0\leq \gamma \leq 0.0635994$, for any $n \in \mathbb{N}$. Further, for sufficiently large value of $n$, the witness operator $W_{n}$  detect the state described by the density operator $\rho_i(\gamma),~~1\leq i \leq 5$ as the matrix realignment criteria
in the same range of $\gamma$.\\
When $0\leq \gamma \leq 0.0635994$, our derived lower bound of the concurrence of the state $\rho_i(\gamma),~~1\leq i \leq 5$ gives better lower bound in comparison to the Albeverio et.al. lower bound of concurrence. This has been shown in Figure-\ref{lambdaimage}.\\
\\

\begin{figure}[h!]
\includegraphics[width=0.45\textwidth]{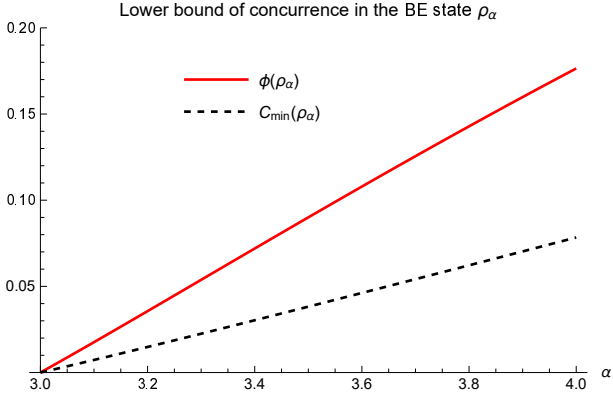}
\caption{The dotted curve represents $C_{min}(\rho_{\alpha})$ and the solid red curve represents the limiting value of our bound, i.e., $\phi(\rho_{\alpha})$ in the PPT entangled region $3 < \alpha \leq 4$. Our bound varies from 0 (at $\alpha = 3$) to 0.176 (at $\alpha = 4$) which gives a better estimate of the lower bound of concurrence as compared to $C_{min}(\rho_{\alpha})$ which varies from 0 (at $\alpha = 3$) to 0.078 (at $\alpha = 4$).}
\label{alphaimage}
\end{figure}

\textbf{Example-3:} Let us again recall a class of bipartite quantum state $\rho_{\alpha}$ for $3 < \alpha \leq 4$, which is given by \cite{horodecki1}
\begin{eqnarray}
\rho_{\alpha}=\frac{2}{7}|\psi^{+}\rangle\langle\psi^{+}|+\frac{\alpha}{7}\sigma_{+}+\frac{5-\alpha}{7}\sigma_{-},~~~3 < \alpha \leq 4
\label{pptes2}
\end{eqnarray}
where $|\psi^{+}\rangle=\frac{1}{\sqrt{3}}(|00\rangle+|11\rangle+|22\rangle)$ and
\begin{eqnarray}
\sigma_{+}=\frac{1}{3}(|01\rangle\langle01|+|12\rangle\langle12|+|20\rangle\langle20|)
\label{sigma+}
\end{eqnarray}
\begin{eqnarray}
\sigma_{-}=\frac{1}{3}(|10\rangle\langle10|+|21\rangle\langle21|+|02\rangle\langle02|)
\label{sigma-}
\end{eqnarray}
The state $\rho_{\alpha}$ represents BES when $3<\alpha \leq 4$. One can check that the marginals of $\rho_{\alpha}$ are maximally mixed and $k_{\rho_{\alpha}} > 0$ for $3< \alpha \leq 4$. We note that the BES of this family are not detected by the witness operators $W^o$, which is defined in (\ref{wit2}).\\
To detect the state $\rho_{\alpha}$, let us construct the witness operator $W_{n}$ with the data given by
\begin{eqnarray}
&&k_{\rho_{\alpha}} = -\frac{64}{27} + \frac{9}{7}\left(1 + \frac{5 - \alpha}{21}\right)^3 \left(1 + \frac{\alpha}{21}\right)^3,\nonumber\\&&
Tr[\rho_{\alpha}^{T_B} \rho_{\alpha}^R] = \frac{2}{21},rank(\rho_{\alpha}^{R}) = 9,\nonumber\\&&
\Lambda_{max}(\rho_{\alpha}^{T_B}) = \frac{1}{21} \sqrt{\frac{33}{2} - 5\alpha + 5\alpha^2 + \frac{5}{2} \sqrt{41 - 20\alpha + 4 \alpha^2}},
\nonumber\\&&  \|\rho_{\alpha}^{R}\|_{1} = \frac{1}{21}(19 + 2\sqrt{19 -15\alpha + 3\alpha^{2}}),\nonumber\\&&
\|\rho_{\alpha}^{R}\|_{2} = \sqrt{\frac{73}{441} + \frac{1}{882} (76 - 60\alpha + 12 \alpha^2)}
\end{eqnarray}
The range of the parameter $\alpha$ for which $Tr(W_{(n)} \rho_{\alpha}) < 0,~~n=1~~\textrm{to}~~5$ is given in Table \ref{alphatable}. It shows that as the value of $n$ increases, more and more BES are detected by the witness operator $W_{(n)}$. \\
\begin{table}[h!]
\begin{tabular}{| p{1cm} | p{3cm} |p{3cm} |}
\hline
\multicolumn{3}{|c|}{The state described by $\rho_{\alpha}$ for $3 < \alpha \leq 4$} \\
\hline
$n$ & The range of $\alpha$ for which $Tr(W_{(n)} \rho_{\alpha})<0$ & $\rho_{\alpha}$ detected by the witness operator $W_{(n)}$ \\
\hline
1 &  $3.7 < \alpha \leq 4$  & BES detected by $W_{(1)}$   \\
\hline
2 &  $3.11 < \alpha \leq 4$ & BES detected by $W_{(2)}$  \\
\hline
3 &  $3.01 < \alpha \leq 4$  & BES detected by $W_{(3)}$ \\
\hline
4 &  $3.0025 < \alpha \leq 4$ & BES detected by $W_{(4)}$  \\
\hline
5 &  $3.0004 < \alpha \leq 4$  & BES detected by $W_{(5)}$ \\
\hline
\end{tabular}
\caption{Detection of BES with the witness operator $W_{n}$ for different $n$ and in the range $3 < \alpha \leq 4$}
\label{alphatable}
\end{table} \\
The witness operator $W_{n}$ not only detect the BES $\rho_{\alpha}$ but also estimate the lower bound $\Phi_{n}(\rho_{\alpha})$ of the concurrence of $\rho_{\alpha}$ when $3 < \alpha \leq 4$. It can be easily shown that the value of $\Phi_{n}(\rho_{\alpha})$ improves as we increase the value of $n$. Thus, for large $n$, the lower bound of the concurrence $\phi(\rho_{\alpha})$ is given by
\begin{eqnarray}
\phi(\rho_{\alpha})=\frac{-1 + \sqrt{19 - 15\alpha + 3\alpha^2}}{\sqrt{111 - 30\alpha + 6\alpha^2}}
\label{lbcex-3}
\end{eqnarray}
Again, the normalized lower bound $C_{min}(\rho_{\alpha})$ of the concurrence of $\rho_{\alpha}$ can be calculated by the prescription given in \cite{chen}
\begin{eqnarray}
C_{min}(\rho_{\alpha})=
\frac{1}{21}(\sqrt{3\alpha^{2}-15\alpha+19}-1)
\label{maxeig}
\end{eqnarray}
We then compare the value of $\phi(\rho_{\alpha})$ given in (\ref{lbcex-3}) with the Albeverio et.al. lower bound given in (\ref{maxeig}). The comparison is shown in the Figure-\ref{alphaimage} and it can be concluded that $C(\rho_{\alpha})\geq \phi(\rho_{\alpha}) \geq C_{min}(\rho_{\alpha})$  where $3 < \alpha \leq 4$.

\begin{figure}[h!]
\includegraphics[width=0.45\textwidth]{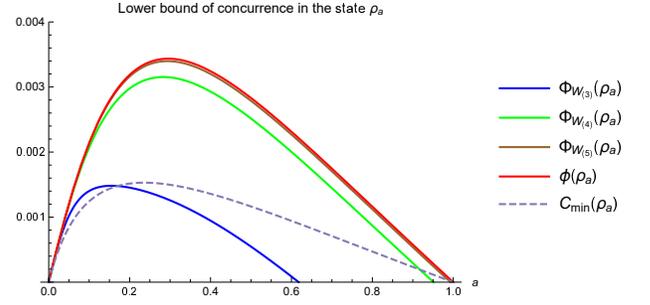}
\caption{Our lower bound $\Phi_{W_{(n)}}(\rho_{a})$, is represented by the solid curves, in blue ($n=3$), green($n=4$), brown ($n=5$). One may note here that for $n>5$, the bound converges to the red curve which represents $\phi(\rho_{a})$. The dotted curve represents $C_{min}(\rho_{a})$, i.e., the normalized lower bound concurrence of $\rho_a$ given by (\ref{nalbebound12}). }
\label{aimage1}
\end{figure}

\textbf{Example-4:} Let us consider another family of BES described by the density operator $\rho_{a}$, which is given by \cite{horodeckia}
\begin{equation}
\rho_a = \frac{1}{8a + 1}
\begin{pmatrix}
a&0&0&0&a&0&0&0&a\\
0&a&0&0&0&0&0&0&0\\
0&0&a&0&0&0&0&0&0\\
0&0&0&a&0&0&0&0&0\\
a&0&0&0&a&0&0&0&a\\
0&0&0&0&0&a&0&0&0\\
0&0&0&0&0&0&\frac{1}{2}(1 + a)&0&\frac{1}{2}\sqrt{1 - a^2}\\
0&0&0&0&0&0&0&a&0\\
a&0&0&0&a&0&\frac{1}{2}\sqrt{1 - a^2}&0&\frac{1}{2}(1 + a)\\
\end{pmatrix}
\end{equation}
where $0\leq a \leq 1$. The state $\rho_a$ is separable when $a = 0$ and 1.\\
It is known that the density matrix $\rho_a$ represents a family of BES for $0 < a < 1$ \cite{horodeckia}. Further, it may be noted that the eigenvalues of the realigned matrix $\rho_a^{R}$ of the state $\rho_a$ and that of the partial transpose of the realigned matrix of $\rho_a$, i.e.,$(\rho_a^R)^{T_B}$, are real and non-negative.\\
In Table-\ref{atable}, we have shown that there exist different witness operators from the family of witness operators $W_{n}$ that can detect BES from the family of states described by the density operator $\rho_{a}$ for different ranges of the parameter $a$.
\begin{table}[h!]
\begin{tabular}{| p{1cm} | p{3cm} | p{3cm} |}
\hline
\multicolumn{3}{|c|}{The state $\rho_{a},~~0<a<1$} \\
\hline
n & The range of the parameter $a$ for which $Tr(W_{(n)}\rho_{a})<0$ &$\rho_{a}$ detected by the witness operator $W_{(n)}$ \\
\hline
1 &  Does not exist  &  $\rho_{a}$ is not detected by $W_{(1)}$ \\
\hline
2 &  $0 < a < 0.016$  &  BES detected by $W_{(2)}$\\
\hline
3 &  $0 < a < 0.62$  &  BES detected by $W_{(3)}$\\
\hline
4 &  $0 < a < 0.951$  &  BES detected by $W_{(4)}$\\
\hline
5 &  $0 < a < 0.9932$  &  BES detected by $W_{(5)}$\\
\hline
6 &  $0 < a < 0.999$  &  BES detected by $W_{(6)}$\\
\hline
7 &  $0 < a < 0.99987$  &  BES detected by $W_{(7)}$\\
\hline
8 &  $0 < a < 0.99998$  & BES detected by $W_{(8)}$\\
\hline
\end{tabular}
\caption{Detection of BES in the range $0 < a < 1$}
\label{atable}
\end{table} \\
Next, our task is to estimate the lower bound of the concurrence of the state $\rho_{a},~~0<a<1$. We first calculate the lower bound $\Phi_{W_{(n)}}(\rho_{a})$ of the concurrence and then compare it with the lower bound $C_{min}(\rho_{a})$ given in (\ref{nalbebound12}). The comparison is shown in Figure-\ref{aimage1} and we find that there exist a critical value of $n$, say $n_{1}$ such that for $n\geq n_{1}$, the quantity $\Phi_{W_{(n)}}(\rho_{a})$ gives better lower bound of the concurrence than $C_{min}(\rho_{a})$. Also, from $Corollary-3$, we know that $\Phi_{W_{(n)}}(\rho_{a}) \rightarrow \phi(\rho_{a}),~~\textrm{as}~~n\rightarrow \infty$, so we obtained the inequality for the state $\rho_{a},~~0<a<1$
\begin{eqnarray}
C(\rho_{a})\geq \phi(\rho_{a}) \geq C_{min}(\rho_{a})
\end{eqnarray} \\

\textbf{Example-5:} Let us consider a $4\otimes4$ dimensional BES $\rho^{(p,q)}$ given by \cite{akbari}\\
\begin{eqnarray}
\rho^{(p,q)} = p \sum_{i=1}^4 |\omega_i\rangle \langle\omega_i| +  q \sum_{i=5}^6 |\omega_i\rangle \langle\omega_i|
\end{eqnarray} where $p$ and $q$ are non-negative real numbers and $4p + 2q = 1$. The pure states $\{|\omega_i\rangle\}_{i=1}^6$ are defined as follows:
\begin{eqnarray*}
&& |\omega_1\rangle = \frac{1}{\sqrt{2}} (|01\rangle + |23\rangle)\\
&& |\omega_2\rangle = \frac{1}{\sqrt{2}} (|10\rangle + |32\rangle)\\
&& |\omega_3\rangle = \frac{1}{\sqrt{2}} (|11\rangle + |22\rangle)\\
&& |\omega_4\rangle = \frac{1}{\sqrt{2}} (|00\rangle - |33\rangle)\\
&& |\omega_5\rangle = \frac{1}{2} (|03\rangle + |12\rangle) + \frac{|21\rangle }{\sqrt{2}}\\
&& |\omega_6\rangle = \frac{1}{2} (-|03\rangle + |12\rangle)+ \frac{|30\rangle }{\sqrt{2}}
\end{eqnarray*}

The state $\rho^{(p,q)}$ becomes invariant under partial transposition when $p = \frac{q}{\sqrt2}$ which implies that $\rho^{(p,q)}$ is a PPT state for $q = \frac{\sqrt{2} - 1}{2}\equiv q_{1}$ and $p = \frac{1 - 2q}{4}\equiv p_{1}$. Since $\|(\rho^{(p_{1},q_{1})})^R\|_1 = 1.08579$, which is greater than 1, so by matrix realignment criteria one can say that $\rho^{(p_{1},q_{1})}$ is a PPT entangled state. Note that for this BES, the realigned matrix $(\rho^{(p_{1},q_{1})})^R$ is hermitian and so is $((\rho^{(p_{1},q_{1})})^R)^{T_B}$. The witness operator $W^o$ defined in (\ref{wit2}) fails to detect this state.
A simple calculation shows that $\rho^{(p_{1},q_{1})}$ is detected by our witness operator $W_{(n)}$ for all $n$, i.e.,
\begin{eqnarray}
Tr(W_{(n)} \rho^{(p_{1},q_{1})}) &=& G< 0, \quad \forall \; n
\end{eqnarray}
where $G= \frac{4}{3} \left(\begin{multlined}\left(\frac{3}{2} - \frac{1}{\sqrt{2}}\right)\left(\frac{7}{128} (-17 + 13\sqrt{2})\right)^n  \\+ \frac{1}{8}(-2 + \sqrt{2})\end{multlined} \right)$.\\
Let us see now how efficiently, we estimate the lower bound of concurrence with the witness operator $W_{n}$. The lower bound $\Phi_{W_{(n)}} (\rho^{(p_{1},q_{1})})$ of the concurrence of the state $\rho^{(p_{1},q_{1})}$ is estimated in Table-\ref{4by4_table}.\\
\begin{table}[h]
\begin{tabular}{| p{1cm} | p{3cm} | p{3cm} |}
\hline
\multicolumn{3}{|c|}{$4\otimes4$ bound entangled state $\rho^{(p_{1},q_{1})}$} \\
\hline
n & $\Phi_{W_{(n)}} (\rho)$ & $C_{min}(\rho^{(p_{1},q_{1})})$ \\
\hline
1 &  0.01757 & 0.0285955 \\
2 &  0.0915681  &  0.0285955 \\
3 &  0.0971719  &  0.0285955 \\
4 &  0.0975963  &  0.0285955 \\
5 &  0.0976284  &  0.0285955\\
\hline
\end{tabular}
\caption{Comparison of $C_{min}$ with our lower bound of concurrence $\Phi_{W_{(n)}}$, for $n = 1$ to 5.}
\label{4by4_table}
\end{table}\\
We then verified the following relation, for $n \geq 1$
\begin{eqnarray}
&&C(\rho^{(p_{1},q_{1})}) \geq \Phi_{W_{(n)}} (\rho^{(p_{1},q_{1})}) \geq C_{min}(\rho^{(p_{1},q_{1})})
\end{eqnarray}
For sufficiently large value of $n$, the lower bound of the concurrence of the state $\rho^{(p_{1},q_{1})}$ is $0.0976311$, which is again much better than $C_{min}(\rho)$. \\

\section{Conclusion}
To summarize, we have constructed different witness operators to detect NPTES and PPTES. Our main contribution in this work is that the constructed witness operator is used to improve the lower bound of the concurrence for any arbitrary entangled state in $d_{1} \otimes d_{2} (d_{1}\leq d_{2})$ dimensional system. In particular, our result will be useful to estimate the lower bound of the concurrence of the BES in higher dimensional systems. This study is important because the exact expression of the concurrence for higher dimensional entangled states is not known and thus one has to depend on the lower bound of it. Also, the witness operator $W_{(n)}$ defined in (\ref{witn1}) is proved to be very useful in detecting not only the NPTES but also in the detection of many BES. Here we have illustrated the above facts with a few examples but one may use $W_{(n)}$ to detect other BES and their lower bound of the concurrence can also be estimated.

\section{Acknowledgements}
The first author Shruti Aggarwal would like to acknowledge Council of Scientific and Industrial Research (CSIR), Government of India, for the financial support in the form of research fellowship (File no. 08/133(0043)/2019-EMR-1).


\begin{thebibliography}{90}
\bibitem{schrodinger} E. Schrodinger, Proceedings of the Cambridge Philosophical Society \textbf{31}, 555 (1935); A. Einstein, B. Podolsky and N.
Rosen, Phys. Rev. \textbf{47}, 777 (1935).
\bibitem{bennett} C. H. Bennett, G. Brassard, C. Creapeau, R. Jozsa, A. Pares and W. K. Wooters, Phys. Rev. Lett. \textbf{70}, 1895 (1993).
\bibitem{bennett1} C. H. Bennett and S. Wiesner, Phys. Rev. Lett. \textbf{69}, 433 (1992).
\bibitem{akpati} A. K. Pati, Phys. Rev. A. \textbf{63}, 014320-1 (2001); C. H. Bennett, D. P. DiVincenzo, P. W. Shor, J. A. Smolin, B. M. Terhal and
W. K. Wooters, Phys. Rev. Lett. \textbf{87}, 077902 (2001).
\bibitem{gisin} N. Gisin, G. Ribordy, W. Tittel and H. Zbinden, Rev. Mod. Phys. \textbf{74}, 145 (2002).
\bibitem{ekert} A. Ekert, and R. Jozsa, Phil. Trans. R. Soc. Lond. A \textbf{356}, 1769 (1998).
\bibitem{rhorodecki} R. Horodecki, P. Horodecki, M. Horodecki, and K. Horodecki, Rev. Mod. Phys. \textbf{81}, 865 (2009).
\bibitem{peres} A. Peres, Phys. Rev. Lett. \textbf{76}, 1413 (1996).
\bibitem{mhorodecki} M. Horodecki, P. Horodecki, and R. Horodecki, Phys. Lett. A \textbf{223}, 1 (1996).
\bibitem{phorodecki} P. Horodecki, J. A. Smolin, B. M. Terhal, and A. V. Thapliyal, Theor. Comput. Sci. \textbf{292}, 589 (2003).
\bibitem{chen} L. Chen and D. Z. Dokovic, J. Phys. A \textbf{44}, 285303 (2011).
\bibitem{divincenzo} D. P. DiVincenzo, P. W. Shor, J. A. Smolin, B. M. Terhal, and A. V. Thapliyal, Phys. Rev. A \textbf{61}, 062312 (2000).
\bibitem{rudolph} O. Rudolph, Quant. Inf. Proc. \textbf{4}, 219 (2005); K. Chen, L.-A. Wu, Quant. Inf. Comp. \textbf{3}, 193 (2003).
\bibitem{mlewenstein} M. Lewenstein, and B. Krauss, J. I. Cirac and P. Horodecki, Phys. Rev. A \textbf{62}, 052310 (2000).
\bibitem{terhal1} B. M. Terhal, arxiv:quant-ph/9810091.
\bibitem{ganguly} N. Ganguly, and S. Adhikari, Phys. Rev. A \textbf{80}, 032331 (2009)
\bibitem{adhikari1} S. Adhikari, N. Ganguly, and A. S. Majumdar,  Phys. Rev. A \textbf{86}, 032315 (2012).
\bibitem{sarbicki} G. Sarbicki, G. Scala, and D. Chruscinski, Phys. Rev. A \textbf{101}, 012341 (2020).
\bibitem{halder} S. Halder, R. Sengupta, Phys. Lett. A \textbf{383}, 2004 (2019).
\bibitem{wiesniak} M. Wiesniak, P. Pandya, O. Sakarya, and B. Woloncewicz, Quan. Rep. \textbf{2}, 49 (2020).
\bibitem{ganguly1} N. Ganguly, S. Adhikari, A. S. Majumdar, J. Chatterjee, Phys. Rev. Lett. \textbf{107}, 270501(2011).
\bibitem{adhikari2} S. Adhikari, N. Ganguly, A. S. Majumdar, Phys. Rev. A \textbf{86}, 032315 (2012)
\bibitem{brandao} F. G. S. L. Brandao, Phys. Rev. A \textbf{72}, 022310 (2005).
\bibitem{mintert2} F. Mintert, Phys. Rev. A \textbf{75}, 052302 (2007).
\bibitem{kchen} K. Chen, S. Albeverio, and S-M Fei, Phys. Rev. Lett.\textbf{95}, 040504 (2005).
\bibitem{phorodecki1} P. Horodecki, R. Augusiak, Quant. Inf. Proc. \textbf{199}, 19 (2006).
\bibitem{gerjuoy} E. Gerjuoy, Phys. Rev. A \textbf{67}, 052308 (2003).
\bibitem{lozinski} A. Lozinski, A. Buchleitner, K. Zyczkowski and T. Wellens, Europhys. Lett. \textbf{62}, 168 (2003).
\bibitem{fmintert} F. Mintert, M. Kus, and A. Buchleitner, Phys. Rev. Lett. \textbf{92}, 167902 (2004).
\bibitem{zou} L. Zou, Y. Ziang, Linear Algebra Appl. \textbf{433}, 1203 (2010).
\bibitem{zhan} X. Zhan, "Matrix Inequalities" (Lecture notes in Mathematics), springer, Berlin, \textbf{1790} (2002).
\bibitem{lin} M. Lin, Sanghai, Czech. Math. J. 66, 737 (2016); P. Zhang, Lin. Alg. and its Appl. \textbf{576}, 258 (2019).
\bibitem{horn} R. A. Horn and C.R. Johnson, Matrix Analysis, Cambridge University Press, USA, 2nd edition, ISBN:0521548233 (2012).
\bibitem{zhao} M-J Zhao, Z-G Li, S-M Fei and Z-X Wang, J. Phys. A: Math. Theor. \textbf{43} 275203 (2010).
\bibitem{anu} A. Kumari, S. Adhikari, Phys. Rev. A \textbf{100},052323, (2019)
\bibitem{bender} C. M. Bender, D. C. Brody, and H. F. Jones, Phys. Rev. Lett. \textbf{89}, 27 (2002).
\bibitem{bender05} C. M. Bender, Contemporary Phys. \textbf{46}, 4, 277-292, (2005).
\bibitem{pati} A K Pati, Pramana J. Phys. \textbf{73}, 3, 485-498, (2009)
\bibitem{bihalan} B. Bhattacharya, S. Goswami, R. Mundra, N. Ganguly, I. Chakrabarty, S. Bhattacharya, A. S. Majumdar, arxiv:quant-ph/2008.12971.
\bibitem{chuang} M.A. Nielson and I. L. Chuang, Quantum Computation and Quantum Information, 10th edition, Cambridge University Press, ISBN 978-1-107-00217-3
\bibitem{horodecki1} P. Horodecki, M. Horodecki and R. Horodecki, Phys. Rev. Lett. 82, 1056 (1999).
\bibitem{cbennett} C. H. Bennett, D. P. DiVincenzo, T. Mor, P. W. Shor, J. A. Smolin, and B. M. Terhal, Phys. Rev. Lett \textbf{82}, 5385 (1999).
\bibitem{bandyo} S. Bandyopadhyay, S. Ghosh and V. Roychowdhury Phys. Rev. A \textbf{71}, 012316 (2005).
\bibitem{horodeckia} P. Horodecki, Phys. Lett. A \textbf{232}(1997), 333-339.
\bibitem{akbari} Y Akbari-Kourbolagh and M. Azhdarghalam, Phys. Rev. A \textbf{99}, 012304 (2019).
\end{thebibliography}
\end{document}